\documentclass[10pt]{article}

\usepackage[hmargin=3cm,vmargin=3cm]{geometry}
\usepackage{graphicx}
\usepackage{amsmath}
\usepackage{rotating}

\begin{document}

\title{Exploring the effect of spatial scales in studying urban mobility pattern}

\author{Hoai Nguyen Huynh\\\\
\textit{Institute of High Performance Computing (IHPC),}\\
\textit{Agency for Science, Technology and Research (A*STAR),}\\
\textit{1 Fusionopolis Way, \#16-16 Connexis, Singapore 138632,}\\
\textit{Republic of Singapore}\\\\
Email: huynhhn@ihpc.a-star.edu.sg}

\date{}

\maketitle

\begin{abstract}
Urban mobility plays a crucial role in the functioning of cities, influencing
economic activity, accessibility, and quality of life. However, the
effectiveness of analytical models in understanding urban mobility patterns can
be significantly affected by the spatial scales employed in the analysis. This
paper explores the impact of spatial scales on the performance of the gravity
model in explaining urban mobility patterns using public transport flow data in
Singapore. The model is evaluated across multiple spatial scales of origin and
destination locations, ranging from individual bus stops and train stations to
broader regional aggregations. Results indicate the existence of an optimal
intermediate spatial scale at which the gravity model performs best. At the
finest scale, where individual transport nodes are considered, the model
exhibits poor performance due to noisy and highly variable travel patterns.
Conversely, at larger scales, model performance also suffers as over-aggregation
of transport nodes results in excessive generalisation which obscures the
underlying mobility dynamics. Furthermore, distance-based spatial aggregation of
transport nodes proves to outperform administrative boundary-based aggregation,
suggesting that actual urban organisation and movement patterns may not
necessarily align with imposed administrative divisions. These insights
highlight the importance of selecting appropriate spatial scales in mobility
analysis and urban modelling in general, offering valuable guidance for urban
and transport planning efforts aimed at enhancing mobility in complex urban
environments.

\textit{Keywords:} Gravity model; Public transport flow; Spatial scales;
Urban mobility pattern; Urban modelling
\end{abstract}

\section{Introduction}

Urban mobility plays a crucial role in shaping the functionality and efficiency
of cities, influencing economic activity, social interactions, and quality of
life. An effective transportation system supports accessibility, reduces
congestion, and enhances urban sustainability \cite{2008@Banister}. Public
transport, in particular, serves as a backbone of mobility in dense urban
environments like Singapore, where land constraints necessitate efficient
transport planning. Understanding travel patterns within the public transport
network is essential for optimising infrastructure, improving service provision,
and informing urban development policies \cite{2024@Zhao.etal}. Analysing these
patterns requires robust models that can capture the complex dynamics of urban
mobility and provide insights into how people move across different spatial
scales \cite{2012@Simini.etal,2018@Toch.etal}.

A variety of models have been developed to study urban mobility, ranging from
agent-based simulations to network-based approaches
\cite{2022@Alessandretti.etal,2017@Anda.etal}. Among them, the gravity model has
been widely used due to its simplicity and effectiveness in capturing aggregate
travel flows \cite{2020@Li.etal}. Inspired by Newton's law of gravity, it
assumes that the interaction between locations is proportional to their
population or activity levels and inversely related to the distance between them
\cite{2018@Barbosa.etal}. The gravity model has been successfully applied in
various urban contexts to estimate mobility patterns and forecast transport
demand \cite{2023@Mepparambath.etal}. However, while the model provides a useful
approximation of mobility flows, its accuracy is arguably influenced by the
spatial scale at which it is applied.

Spatial aggregation has been shown to play a critical role in urban mobility
modelling, affecting both data representation and model performance
\cite{2023@BinAsad.Yuan,2019@Dabiri.Blaschke,2022@Oshan.etal}. At fine spatial
scales, such as individual bus stops or train stations, the models may struggle
to capture meaningful patterns due to high variability and noise in travel
behaviour, leading to overfitting or poor generalisability. Conversely, at very
coarse spatial scales, over-aggregation may lead to a loss of critical mobility
details, obscuring important underlying dynamics and interaction patterns, and
resulting in model underperformance. These challenges reflect the Modifiable
Areal Unit Problem (MAUP), a well-known issue in spatial analysis where
different zoning schemes or levels of aggregation can lead to significantly
different analytical results (sometimes referred to as ``Openshaw effect''
\cite{2022@Goodchild}). In mobility research, this means that both the
resolution and method of spatial aggregation must be carefully chosen to ensure
accurate model interpretation and policy relevance.

Recent studies have sought to quantify and mitigate the effects of spatial scale
in mobility modelling. For instance, it has been showed that while some mobility
metrics (e.g. radius of gyration and entropy) remain relatively stable across
scales, others vary significantly depending on the spatial resolution,
influencing how individual activity spaces are characterised
\cite{2024@BinAsad.Yuan,2020@Wang.Yuan}. Similarly, spatial boundaries can be
argued to critically affect the predictive power of mobility models
\cite{2008@Rozenfeld.etal,2012@Simini.etal}, which means that spatial scale may
not be merely a technical detail but a foundational element of mobility theory.
However, while it has been acknowledged that model performance can vary
significantly depending on the chosen spatial resolution, there is limited
consensus on the optimal level of aggregation \cite{2019@Chen.etal,2023@Huynh}.
Furthermore, spatial aggregation based on administrative boundaries may not
necessarily align with actual urban movement patterns, potentially introducing
biases in mobility analysis.

Despite extensive research on urban mobility modelling, several gaps remain in
understanding the interaction between spatial scale and model performance using
public transport flow data. First, the impact of spatial scale on gravity model
performance has not been systematically explored in the context of public
transport networks, particularly in highly urbanised environments like
Singapore. Second, while administrative boundaries are often used for spatial
aggregation, their effectiveness compared to alternative aggregation methods
like distance-based clustering remains unclear. Moreover, most of these studies
have focused on either individual-level GPS \cite{2020@Alessandretti.etal},
mobile phone records \cite{2015@Louail.etal} or social media data
\cite{2024@BinAsad.Yuan}, rather than formal public transport usage data, which
reflects structured and policy-relevant travel behaviour. This study addresses
these gaps by examining how spatial aggregation affects gravity model
performance using public transport flow data in Singapore. Specifically, the
influence of different spatial scales on model accuracy is investigated, and
comparison between administrative boundary-based aggregation and distance-based
methods is made to evaluate which better captures urban mobility patterns. The
findings from this study can provide insights into optimal spatial resolutions
for mobility analysis and inform urban and transport planning strategies.

The remainder of this paper is organised as follows. Section 2 describes the
data and methods used in the study, including details on public transport flow
data, spatial aggregation approaches, and gravity model fitting procedures.
Section 3 presents the results and discussion, focusing on model performance
across different spatial scales and aggregation methods, as well as comparison
of mobility pattern between time windows. Finally, Section 4 concludes with key
findings, implications for urban planning, and potential directions for future
research.

\section{Data and methods}

\subsection{Data}
\label{sec:data}

The datasets used in this study were obtained from relevant authorities in
Singapore, and can be categorised by public transport and administrative
boundaries.

The public transport related data was obtained from the Land Transport Authority
(LTA) of Singapore, which provides comprehensive data on public transportation
infrastructure and usage across the city-state \cite{LTA_DataMall}. The data
contains the location of bus stops and train stations and the amount of traffic
flow between them. As this study focuses on mobility pattern within Singapore,
the bus stops in Johor (Malaysia) that are parts of the cross-border services
between Singapore and Malaysia are excluded. The traffic flow data contains
information on the number of trips made between a pair of origin and destination
transport nodes during hourly time windows (from \texttt{HH:00} to
\texttt{HH:59}) that will be merged to obtain the daily flow on a typical day of
a month, for both weekdays and weekends. The public transport flow data used in
this study is for October 2024, which was chosen to reflects recent typical
mobility patterns without the anomalies and seasonal variations in travel
behaviours during major holiday periods in Singapore.

In addition to transport flow data, administrative boundaries delineated in the
Master Plan 2019 \cite{Data_Gov} by the Urban Redevelopment Authority (URA) are
also used to compute different levels of spatial aggregation of transport nodes.
These boundaries include three hierarchical levels: subzone, planning area, and
region. Subzones represent the most granular administrative divisions in
Singapore, while planning areas and regions provide broader spatial groupings.

\subsection{Spatial clustering of nodes}

\begin{figure}[t]
\centering
\includegraphics[width=\textwidth]{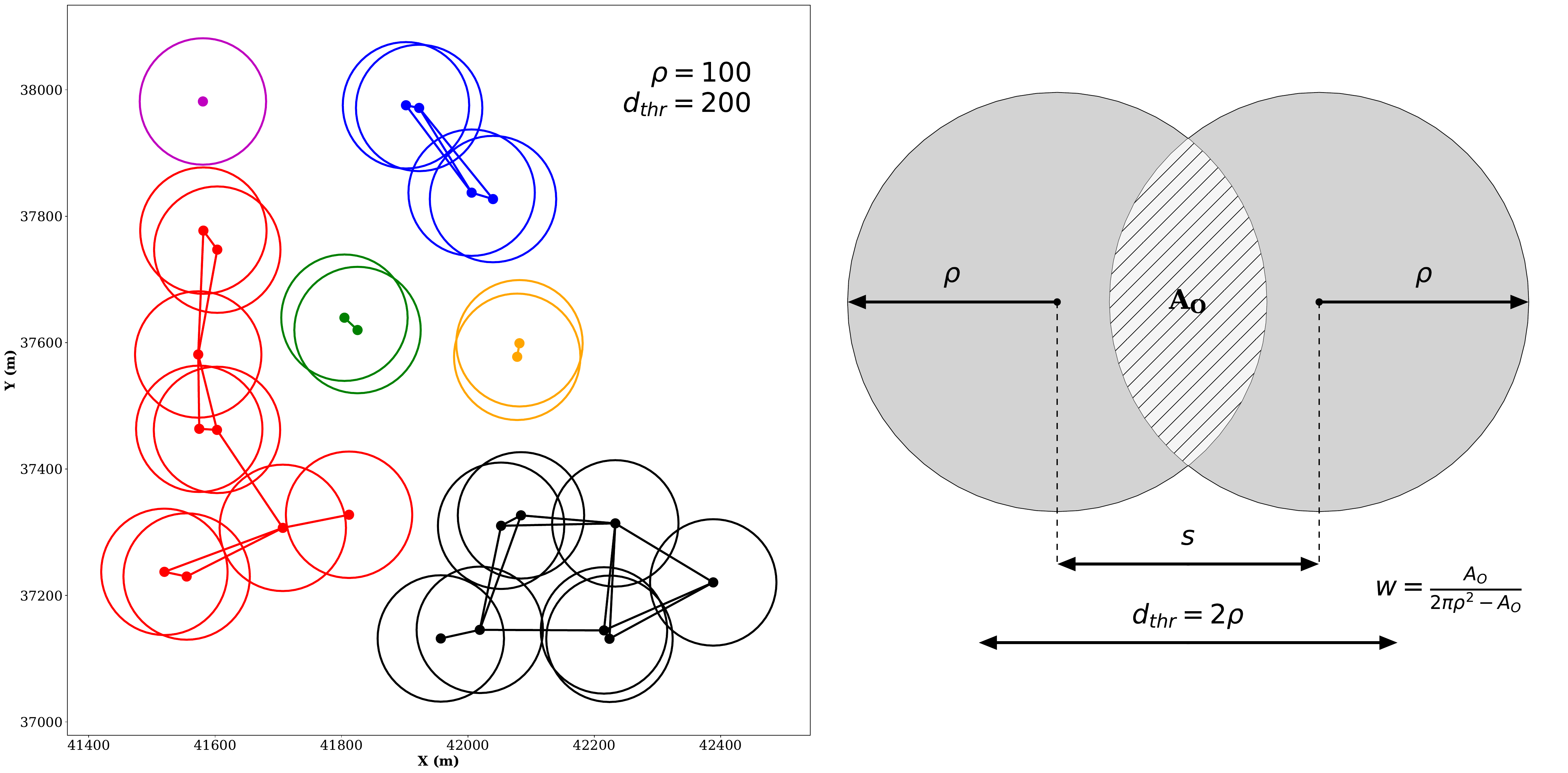}
\caption{Construction of network of transport nodes in which links and
corresponding weights $w$ are determined by the overlap area (hatched) of buffer
circles of radius $\rho$ centred at the nodes (right panel). The threshold
distance for a pair of nodes to be considered in the same cluster is
$d_{thr}=2\rho$, beyond which the buffer circles do not overlap (left panel).
\label{fig:network}}
\end{figure}

Apart from the spatial aggregation by administrative boundaries, the transport
nodes can also be clustered by spatial proximity. In this study, a procedure is
devised to identify the clusters of transport nodes given a distance parameter.
First, a network is contructed for all public transport nodes in Singapore with
links added between pairs of nodes whose Euclidean distance is smaller than a
given threshold $d_{thr}$. The weight of such links is calculated as the ratio
between the overlap area of the buffer circles of radius $\rho=d_{thr}/2$
centred at the nodes and their union area (see Fig. \ref{fig:network}). This
area ratio reflects the strength of relationship between two nodes in terms of
how close they are to one another. The network will then be divided into
clusters using a procedure of community detection based on modularity (similar
to previously employed in \cite{2019b@Huynh}). The clustering procedure is
described in details in \cite{2022@Jiang.Huynh}, involving multiple runs of the
Louvain method for community detection and effective average of clustering
patterns to identify the converged clusters of nodes.

Different levels of spatial clustering are obtained by varying the distance
threshold parameter $d_{thr}$ from 0 to $6,000$ m in steps of $100$ m. For every
value of $d_{thr}$, the Louvain community detection algorithm is applied $100$
times to yield the clusters of nodes. These distance-based clusters together
with administrative boundaries (subzone, planning area and region) will serve as
different kinds of spatial aggregation to assess the performance of the urban
mobility flow model described in the next section.

\subsection{Modelling the urban mobility flow}

The gravity model has been widely used in transportation and urban studies to
predict mobility flows between locations \cite{2023@Mepparambath.etal}. It is
based on the analogy of Newton's law of gravity, where the interaction between
two places is proportional to their population (or activity level) and inversely
related to the distance between them. In the context of urban mobility, the
model estimates the volume of trips between origin and destination locations and
can be expressed as
\begin{equation}
F_{ij} = G\frac{M_i^\alpha M_j^\beta}{D_{ij}^\gamma}\label{eq:gravity_model}
\end{equation}
in which $F_{ij}$ denotes the traffic volume from location $i$ to $j$, $G$ is
some scaling constant, $M_i$ and $M_j$ denote the corresponding activity level
at these locations, and $D_{ij}$ the distance between them, whereas $\alpha$,
$\beta$, and $\gamma$ are associated parameters to be fitted using the mobility
data. In this study, the total outflow traffic at location $i$ and inflow
traffic at location $j$ are used as proxy for their activity level. The distance
between the locations is taken as the Euclidean distance between the centroid of
the cluster of transport nodes. It should be noted that in the case of
administrative boundaries, the centroid is not the centroid of the polygon but
the centroid of the cluster of transport nodes contained within the polygon.

The gravity model is then fitted using linear regression, where the logarithm of
observed mobility flows is modelled as a function of explanatory variables
including the activity level at origin and destination locations and the
distance between them. This is achieved by employing the logarithmic form of Eq.
\ref{eq:gravity_model}
\begin{equation}
\log{F_{ij}} = \omega + \alpha\log{M_i} + \beta\log{M_j} - \gamma\log{D_{ij}}
\label{eq:gravity_model_log}
\end{equation}
in which the model parameters $\alpha$, $\beta$, and $\gamma$ are estimated
using ordinary least squares (OLS) regression. Goodness-of-fit is evaluated
using the coefficient of determination $R^2$ to assess how well the model
explains variations in urban mobility flows between locations.

As the number of data points varies with different levels of spatial
aggregation, the adjusted $R^2$ \cite{1997@Raju.etal} is used to characterise
the quality of model fitting instead of the usual $R^2$ to account for the data
size and the complexity of the model (i.e. the number of independent variables).
The formula for adjusted $R^2$ is given by $R_{adj}^2=1-(1-R^2)(n-1)/(n-p-1)$ in
which $n$ is the number of data points and $p$ the number of parameters. Given
the gravity model has been shown to work very well with urban mobility pattern,
aggressive test of the model in this study will be performed by using only $50\%$
of the data for training and the model is tested on the remaining $50\%$.

\section{Results and discussion}

\subsection{Patterns of different levels of spatial aggregation}
\label{sec:result_pattern}

\begin{sidewaysfigure}
\centering
\includegraphics[width=0.91\textwidth]{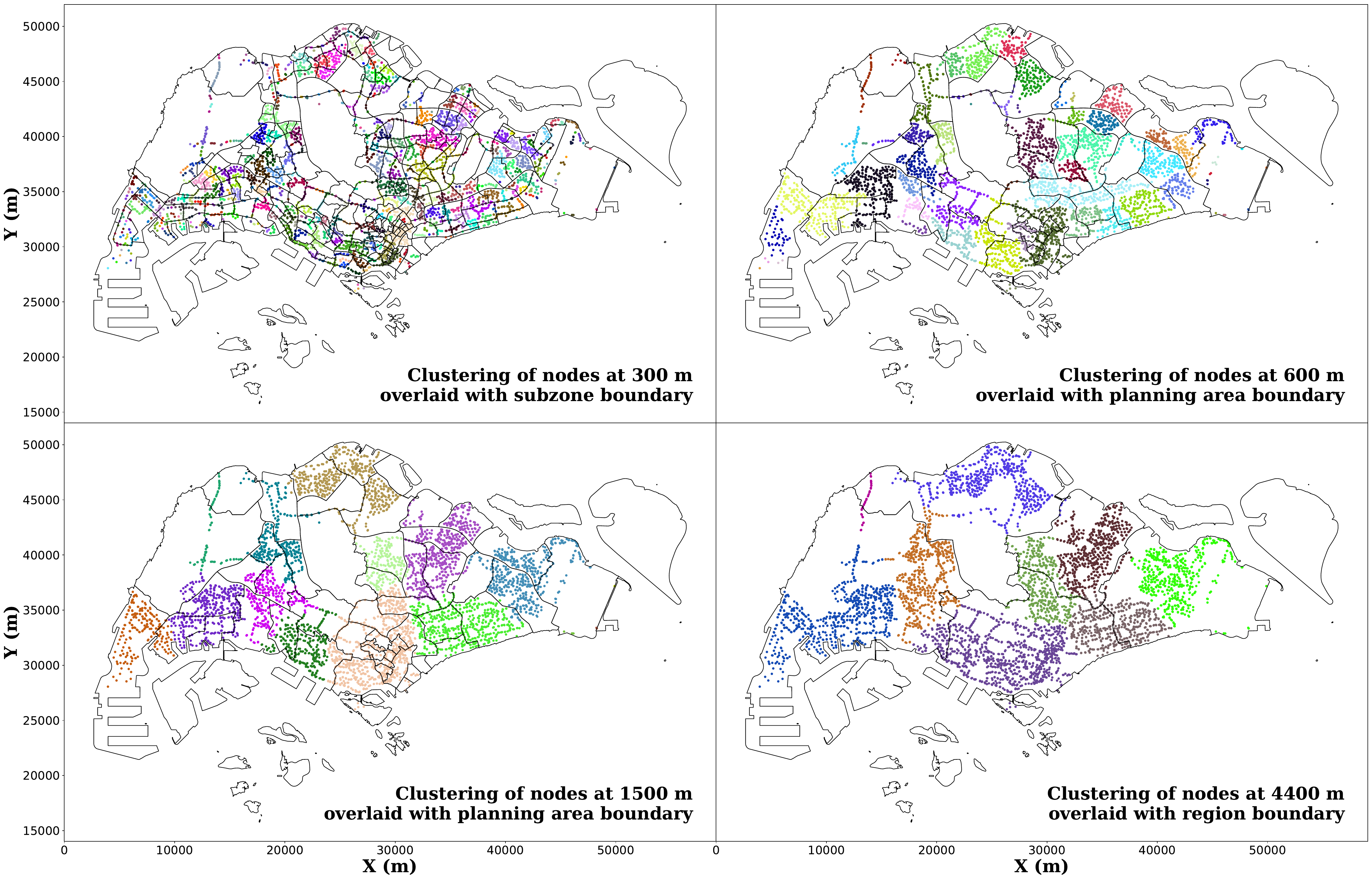}
\caption{Clusters of transport nodes at different values of threshold distance
$d_{thr}$.} \label{fig:clusters}
\end{sidewaysfigure}

As the value of the distance threshold $d_{thr}$ varies, different clustering
patterns of transport nodes are observed. At $d_{thr}=0$, every node forms its
own cluster, whereas the nodes are grouped into $6$ clusters at $d_{thr}=6,000$
m. The clustering patterns of nodes at different values of $d_{thr}$ are shown
in Fig. \ref{fig:clusters}. These patterns are also compared with clustering of
nodes by subzones, planning areas and regions to assess their alignment with
administrative boundaries. It could be observed that the distance-based
aggregation does not necessarily align with imposed administrative boundaries,
suggesting nuanced differences in patterns of spatial organisation across
scales. The selected distance threshold values of $300$ m, $600$ m, and $4,400$
m in Fig. \ref{fig:clusters} show the clustering patterns that most closely
match the clustering by administrative boundaries, as quantified by the mutual
information score, which is commonly used to compare sets of different subset
structures \cite{2010@Nguyen.etal}. The identified clusters at different spatial
aggregration levels will be used to assess the impact of spatial scale on the
performance of the gravity model, providing insights into the relationship
between spatial aggregation and urban mobility dynamics.

\subsection{Performance of gravity model across spatial scales}

\begin{figure}[t]
\includegraphics[width=\textwidth]{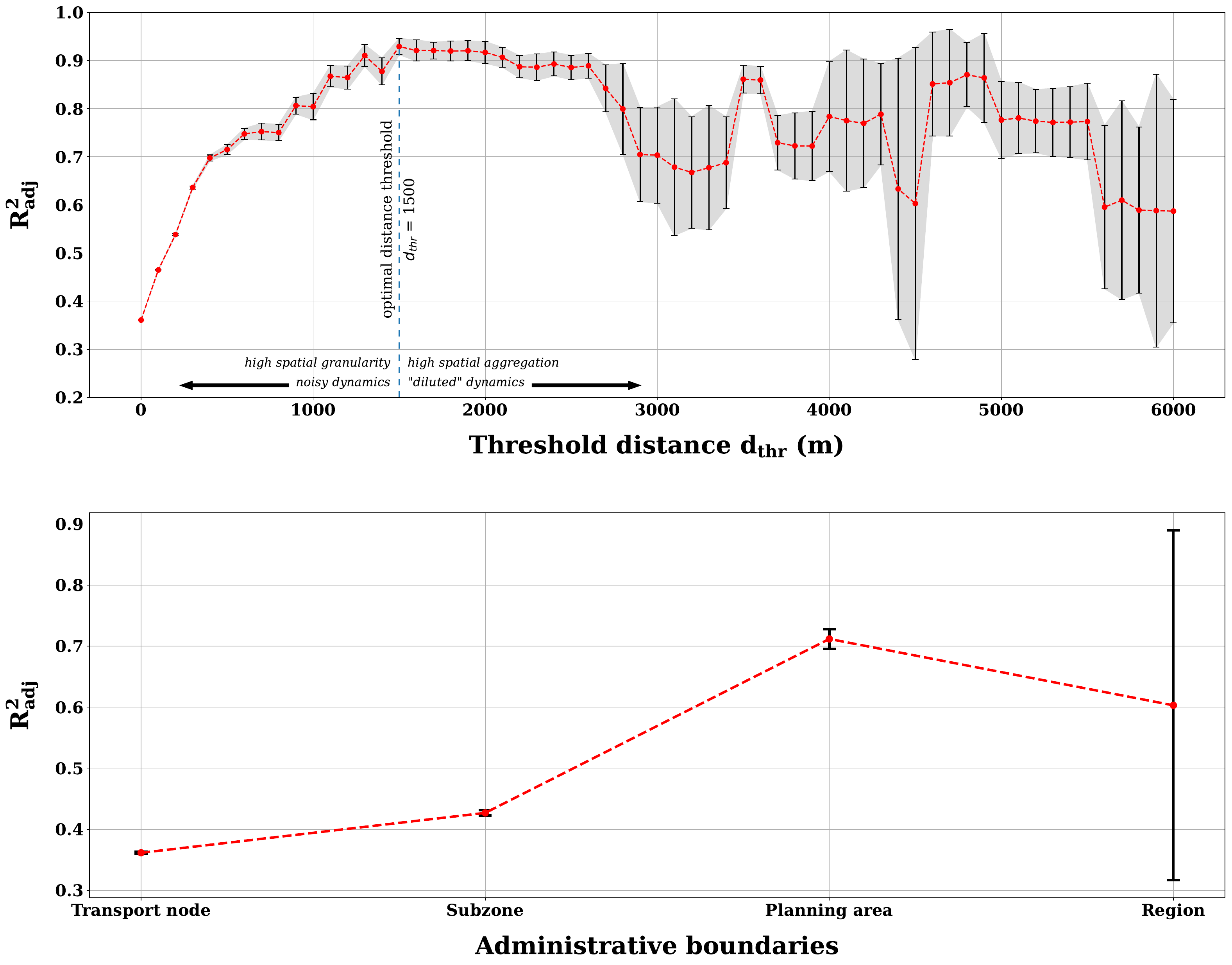}
\caption{Quality of fitting the gravity model to weekday mobility flows at
different levels of spatial aggregation by distance threshold (top) and
administrative boundaries (bottom). At each spatial aggregation, the average
$R_{adj}^2$ value and its error bar are computed over 100 runs with
randomisation of 50:50 train-test split of the data.
\label{fig:result_performance}}
\end{figure}

For every clustering structure of the transport nodes, the gravity model is
fitted to the corresponding aggregated traffic flow pattern to assess its
performance across spatial scales. In order to obtain a reliable measure of the
performance, the model fitting is run 100 times with randomisation of 50:50
train-test split. Equation \ref{eq:gravity_model_log} is fitted using $50\%$ of
the data to estimate the parameters $\alpha$, $\beta$, and $\gamma$, and the
model performance is assessed based on its prediction of the remaining $50\%$ of
the data. The results for weekday mobility pattern (see Fig.
\ref{fig:result_performance}, top panel) show that the fitting at the least
aggregate level, the transport node, is the worst with $R_{adj}^2$ being only
around $0.35$, meaning that less than $40\%$ of variance in the traffic flow can
be explained by the combination of total outflow at origin, total inflow at
destination and the distance between them. The model performance quickly
improves as the nodes become spatially aggregated. The same argument as in
\cite{2023@Huynh} could be made that the aggregation of nodes better reflects
the underlying dynamics of traffic flows where commuters from a particular
location may use multiple transport nodes within the vicinity.

As the spatial aggregation increases, the average quality of fitting reaches the
peak value at $d_{thr}=1,500$ m and starts to decline afterwards. This decline
signals that the transport nodes may be over-aggregated and that further
congregating nodes may indeed ``dilute'' the dynamics of traffic flows whereby
the true pattern is not captured as well as by a smaller mass of nodes, i.e. the
flows are aggregated more than necessary. It is worth pointing out that the
quality of fitting at large spatial scales fluctuates significantly compared to
smaller ones, indicating low reliability of the fitting. Apart from diluting
dynamics, the fact that the number of data points decreases with higher level of
spatial aggregation may also contribute to a poorer fitting of the model when
its complexity is not justified by the amount of data available.

\subsection{Comparison of mobility pattern across temporal windows}

\begin{figure}[t]
\includegraphics[width=\textwidth]{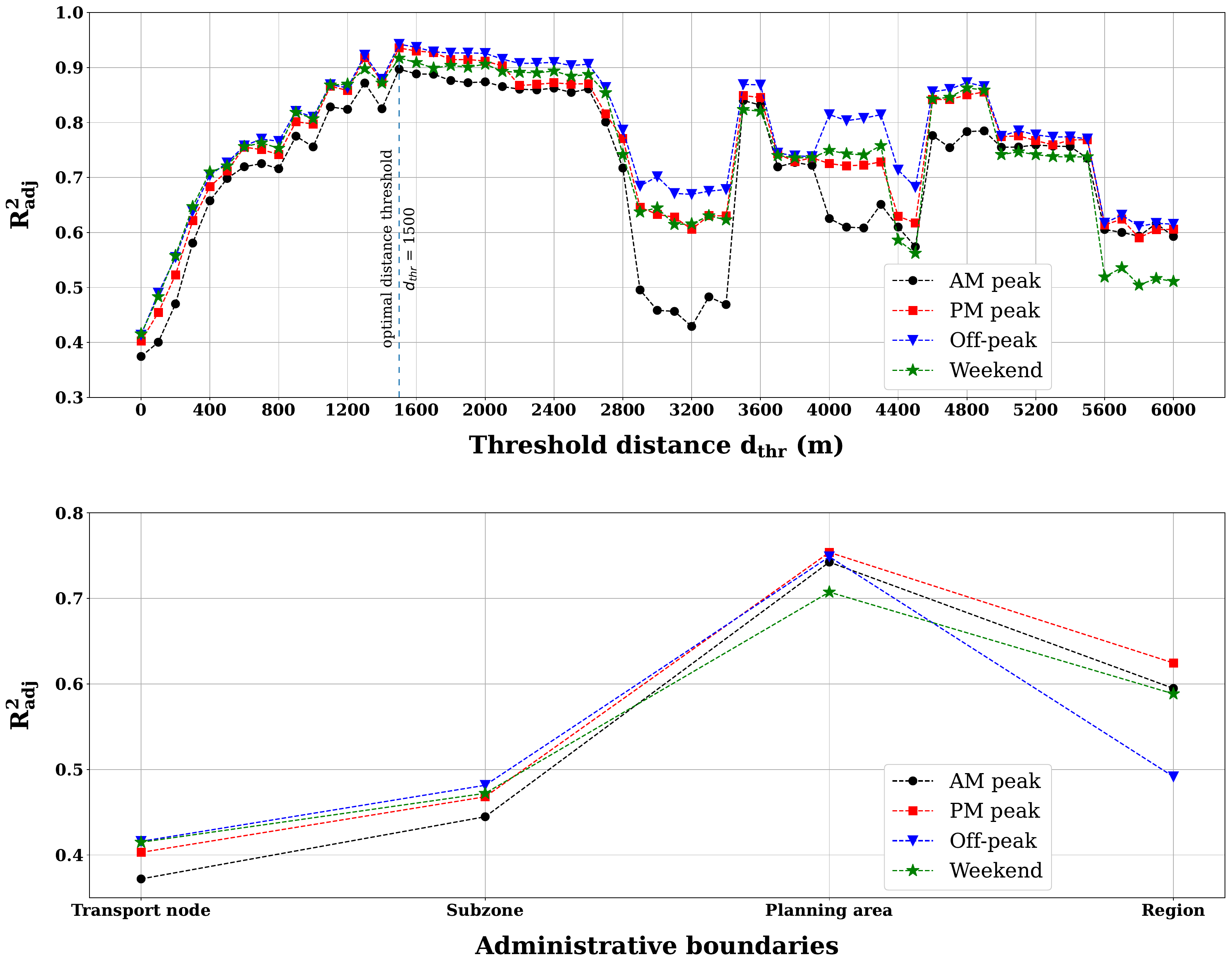}
\caption{Comparison of gravity model fitting by diffent periods: weekday AM
peak, weekday PM peak, weekday off-peak, and weekend. Quality of fitting the
gravity model at different levels of spatial aggregation by distance threshold
(top) and administrative boundaries (bottom).\label{fig:temporal_windows}}
\end{figure}

To further examine the temporal consistency of the gravity model performance, a
stratified analysis is conducted based on time of day and weekday versus
weekend travel patterns. The model is separately fitted to public transport
mobility data from three distinct weekday periods, namely AM peak (6:00 AM to
before 10:00 AM), PM peak (4:00 PM to before 8:00 PM), and off-peak hours (10:00
AM to before 4:00 PM), as well as to aggregated flows over the entire day on
weekends. Across all time windows, the gravity model consistently shows the best
performance when spatial aggregation is applied at 1,500 m (see Fig.
\ref{fig:temporal_windows}, top panel). This suggests a robust spatial scale at
which urban mobility dynamics in Singapore are optimally captured, regardless of
temporal variation in travel behaviour. While minor fluctuations in model
fitting are observed between time periods (likely due to differing trip purposes
and passenger profiles), the spatial scale of 1,500 m provides a stable balance
between granularity and aggregation. These findings reinforce the notion that
intermediate spatial scales can effectively reduce noise in fine-grained data
without oversimplifying travel patterns, making them suitable for both weekday
commuting and weekend leisure mobility analysis.

\subsection{Effect of different spatial aggregation methods}

A similar trend in the quality of fitting the model is also observed using the
administrative boundaries as the method of spatial aggregation (see Fig.
\ref{fig:result_performance}, bottom panel). At the first level of subzone, the
fitting shows some improvement with $R_{adj}^2$ rising above $0.4$. The
improvement continues at planning area level when the average quality of fitting
reaches $0.7$. However the trend does not hold beyond planning area when the
model performs poorer at region level, indicating over-aggregation of transport
nodes. Recalling the corresponding patterns of clusters based on distance
threshold and administrative boundaries in Sec. \ref{sec:result_pattern},
grouping of transport nodes at subzone level is closest to the pattern of
clusters at $300$ m, at planning area level the distance is $600$ m, and region
level maps best to $4,400$ m (see Fig. \ref{fig:clusters}). The corresponding
quality of fitting at these distance also shows comparable values with similar
trend as the spatial aggregation level increases. It should be noted that all of
these administrative boundary-based clusterings perform worse than the spatial
aggregation at $1,500$ m. The poorer performance of spatial aggregation based on
administrative boundaries compared to distance-based aggregation indicates that
these boundaries may not accurately reflect the true patterns of movement and
interaction among transport nodes.

These results are also consistent across temporal windows when fitting mobility
data from different periods of weekdays (AM peak, PM peak, and off-peak) as well
as on weekends (see Fig. \ref{fig:temporal_windows}, bottom panel). Similar to
the results for weekday mobility data, the gravity model exhibits the best
performance at planning area level among the administrative boundaries when
fitting stratified data from these windows, with the average $R_{adj}^2$ value
peaking around $0.7$. Notably, while the whole-day data on weekdays (Fig.
\ref{fig:result_performance}, bottom panel) and weekends (green star marker in
Fig. \ref{fig:temporal_windows}, bottom panel) show similar peak $R_{adj}^2$
value of $0.7$, the sub-weekday periods all show slightly higher peak value.
This hints at the variability in mobility behaviour throughout the day affecting
the model performance. In the same vein, the noticeably worse performance of the
gravity model at region scale for off-peak period (blue triangle marker in Fig.
\ref{fig:temporal_windows}, bottom panel) compared to whole days or peak periods
could be due to irregular large-scale movement pattern when long-distance travel
appears to be rare outside rush hours. Nevertheless, these observations require
further substantiation which is beyond the scope of the current study.

\subsection{Contribution to mobility research and future directions}

This study employs the gravity model as a mean to illustrate the effect of
spatial scales and units in urban modelling. While the gravity model offers a
simple and intuitive framework for modelling urban mobility, it is not without
limitations. It assumes that flows between locations depend solely on size and
distance, overlooking other influential factors such as land use mix, transport
connectivity, and individual travel preferences, which may influence local
variations and dynamic behaviours inherent in real-world mobility. Despite these
limitations, this study contributes to the field by systematically evaluating
how spatial scale affects the model performance, offering practical guidance on
appropriate aggregation levels for transport analysis. Furthermore, it
highlights the mismatch between administrative boundaries and actual mobility
patterns, suggesting that data-driven, distance-based approaches may yield more
accurate representations of urban movement. Future research could build on these
findings by incorporating additional variables into the model, such as
socio-demographic factors or transport service attributes. The role of spatial
scale can also be explored in other modelling approaches like radiation
\cite{2013@Masucci.etal} or machine learning-based models \cite{2018@Toch.etal}
to provide more comprehensive understanding of complex travel behaviours in
urban systems.

Moving toward practical applications of these findings, actionable insights can
be derived by combining mobility patterns with contextual knowledge of the
actual urban organisation. In the case of Singapore, the clustering pattern
observed at the 1,500-m aggregation level (see Fig. \ref{fig:clusters}, bottom left
panel) suggests strong functional integration between areas such as Jurong East
and Bukit Batok (violet cluster around X=16,000 and Y=35,000). Similarly, northern
neighbourhoods like Woodlands, Sembawang, and Yishun, or northeastern areas such
as Serangoon, Hougang, Sengkang, and Punggol, may benefit from being considered
as cohesive planning units. These spatial patterns highlight the potential for
more integrated planning strategies that align with how residents actually move
through the city-state. Further analysis, such as incorporating the spatial
distribution of amenities, residential density, and public transport
infrastructure like bus routes and train stations, could provide deeper insights
into mobility demand and service accessibility. While such integration lies
beyond the scope of the current study, it represents a promising direction for
future research. In other urban contexts, similar approaches could offer
powerful tools for urban planning by combining mobility data with additional
layers of urban organisation, such as land use and the spatial distribution of
services and infrastructure.

\section{Conclusion}
In this study, a computational method is developed to analyse the urban mobility
pattern in Singapore. The findings reveal that while the gravity model can
generally capture the flow dynamics, its performance quality varies
significantly when different spatial units are used to calculate the amount of
traffic between origin and destination locations in the model. It is found that
the model fits poorly at the transport node level and performs best at some
intermediate level of spatial aggregation corresponding to a threshold distance
of $1,500$ m between nodes. Beyond that scale, the model performance decreases,
signaling over-aggregation. Similar pattern is observed if the administrative
boundaries are used, where the model fits poorly at the lowest level of subzone
and improves at the intermediate level of planning area before decreasing at the
highest level of region. However, the spatial aggregation at these
administrative boundaries perform poorer than the distance-based aggregation,
indicating that the administrative boundaries are artificial and not reflective
of the actual organisation of mobility patterns on the ground.

The findings here offer valuable insights into the spatial organisation of urban
areas in Singapore. The method developed in this study could be used to identify
functional urban areas at different scales when combined with other relevant
datasets. Additionally, this approach can help reveal latent mobility patterns
and interactions between different parts of a city, offering a data-driven lens
through which to interpret urban dynamics. Both the methodology and results can
be useful for relevant urban and transport planning authorities in understanding
the impact of physical infrastructure on urban mobility behaviours so that
future land use and transport network can be effectively developed. Future
research could build on this work by incorporating additional layers of spatial
information, such as the distribution of amenities, residential densities, and
the structure of public transport network. This would allow for a more
comprehensive analysis of urban function and accessibility, ultimately
contributing to the development of more inclusive and efficient urban
environments.

\section*{Acknowledgement} This research is supported by A*STAR project number
CoT-H1-2025-3 under the Cities of Tomorrow Grant 2024.

\bibliographystyle{unsrt}
\bibliography{references}

\end{document}